# Target Design for XUV Probing of Radiative Shock Experiments

*U. Chaulagain, C. Stehlé, P. Barroso, M. Kozlova,
J. Nejdl, F. Suzuki Vidal and J. Larour*

**Editors:**

Dr. Binod Adhikari

Dr. Manoj Kumar Yadav

Mr. Kiran Pudasainee

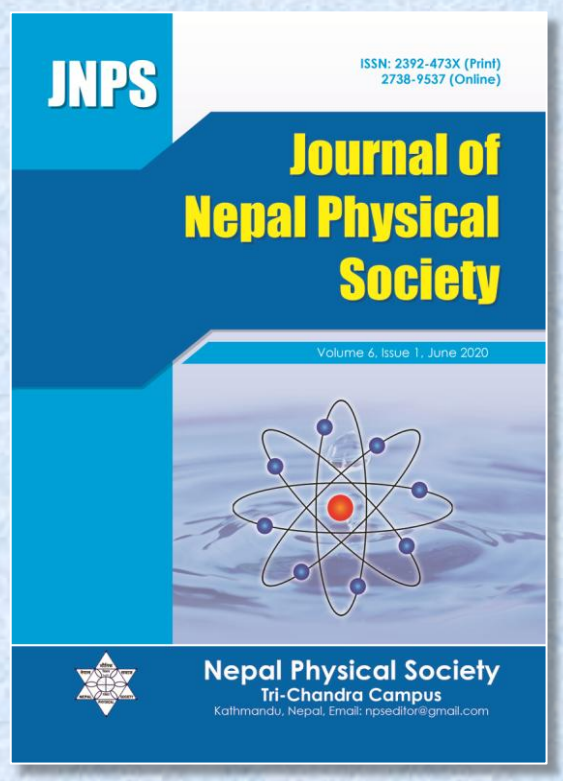

# Target Design for XUV Probing of Radiative Shock Experiments

U. Chaulagain[1, *], C. Stehlé[2], P. Barroso[3], M. Kozlova[1, 4], J. Nejdl[1], F. Suzuki Vidal[5] and J. Larour[6]

[1]ELI Beamlines, Institute of Physics ASCR, Na Slovance 1999/2, Prague, 182 21, Czech Republic
[2]LERMA UMR 8112, Observatoire de Paris, PSL Research University, UPMC, Sorbonne University, CNRS, France
[3]GEPI, UMR 8111, Observatoire de Paris, PSL Research University
[4]Institute of Plasma Physics ASCR, Za Slovankou 1782/3, Prague, 182 00, Czech Republic
[5]The Blackett Laboratory, Imperial College, Prince Consort Road, London SW7 2AZ, UK
[6]LPP, UMR 7648, Ecole Polytechnique, UPMC, Sorbonne Université, U-PSud, 91128 Palaiseau, France

[*]Corresponding Email: uddhab.chaulagain@eli-beams.eu

**Abstract**

Radiative shocks are strong shocks characterized by plasma at a high temperature emitting an important fraction of its energy as radiation. Radiative shocks are commonly found in many astrophysical systems and are templates of radiative hydrodynamic flows, which can be studied experimentally using high-power lasers. This is not only important in the context of laboratory astrophysics but also to benchmark numerical studies. We present details on the design of experiments on radiative shocks in xenon gas performed at the kJ scale PALS laser facility. It includes technical specifications for the tube targets design and numerical studies with the 1-D radiative hydrodynamics code MULTI. Emphasis is given to the technical feasibility of an XUV imaging diagnostic with a 21 nm (~58 eV) probing beam, which allows to probe simultaneously the post-shock and the precursor region ahead of the shock. The novel design of the target together with the improved X-ray optics and XUV source allow to show both the dense post-shock structure and the precursor of the radiative shock.

## 1. INTRODUCTION

Radiative shocks are found in many astrophysical scenarios, e.g. during the stellar breakthrough of a supernova shock [1], in accretion shocks [2, 3], interaction of supernova remnants with the interstellar medium (ISM) [4, 5], and in the stellar jets from young stars [6, 7]. All these shocks are characterized by high Mach numbers (M>>1). For instance, in accretion shocks in young stars, matter is channelled from the protostellar disk towards the photosphere at ~400 kms$^{-1}$ leading to an accretion shock and a subsequent reverse shock propagating backwards. This leads to strong heating (up to several $10^6$ K) and strong X-ray emission [8-13]. In general, the angular resolution of present astronomical observations is not sufficient to resolve the structure of these shocks, and thus their fine structure and dynamical behaviour remains poorly constrained.

Understanding the properties of radiative shocks is a fundamental problem since they are the basis for the interpretation of observations of various astrophysical phenomena. At high shock velocities, a significant fraction of the energy in the shock is converted to radiation. Depending on the density of the upstream medium, radiation can escape freely or be re-absorbed in the medium where the shock propagates, ionizing it and forming a 'radiative precursor'. This process is usually accompanied by strong post-shock cooling [14, 15].

High-power lasers and pulsed power Z-pinches are capable of creating high energy density

*U. Chaulagain, C. Stehlé, P. Barroso, M. Kozlova, J. Nejdl, F. Suzuki Vidal and J. Larour*

environments allowing to study a wide variety of states of matter of astrophysical interest over many orders of magnitude in density, pressure and temperature [16]. Several radiative shock experiments have been performed in different laboratories in the last decades, aiming at understanding the coupling of radiation with hydrodynamics and the role of radiation in the shock structure [17-28]. Measurements of these extreme plasma conditions are critical for theoretical modelling and to validate numerical simulations. Although the conditions for shock generation and characteristic plasma parameters between the laboratory and space can be significantly different.

There are three dimensionless numbers; Mach number (M), Boltzmann number (Bo) and Mihalas number (R), are commonly used to characterized the measure the strength of the shock and the effect of the radiation in the shock. The strength of a shock is defined Mach number (M), which is the ratio of the shock velocity to the local sound speed. The local sound speed can be taken either with the initial gas condition or upstream gas conditions (conditions of Radiative Precursor). Bo and R determines the extents of radiation that plays the role in shock dynamics. Bo compares the local material energy flux to the radiative flux and R compares the internal energy density with radiation density. If Bo ≤ 1, the flow is in the radiation flux dominated regime. The radiative shock we reported is a radiation flux dominated regime shock wave with the value of Bo is ~0.01. Some of the recent experiments have demonstrated the ability to obtain dimensionless Mach and Boltzmann number of the same order of magnitude and thus both phenomena are in conditions where radiative flux plays an important role see e.g. [21].

However, it is important to note that the coupling (e.g. absorption) of radiation with hydrodynamics is more important in the laboratory than in astrophysics (where the medium is relatively dilute) making one to one comparison challenging. In the particular case of stellar accretion shocks [11], e.g. at the interface of an accretion column with the photosphere of a star, previous experiments (see e.g. [20-28] and references therein) have looked at narrowing the gap between astrophysics and the laboratory by producing shocks with velocities of ~10-100s $kms^{-1}$, i.e. upstream Mach numbers ~100 (with respect to cold gas) and Boltzman numbers ~$10^{-2}$-$10^{-3}$.

Although radiative shocks have been extensively studied in the laboratory, the majority of previous experiments have focused either on the study of the radiative precursor using optical laser probing [17, 18], or the post shock structure with hard X-ray backlighting [21, 24]. Although optical probing is often used to characterize laser-produced plasmas, it is limited to moderate electron densities due to strong absorption of the probing laser beam near the critical density (i.e. $6.4 \times 10^{20} cm^{-3}$ at 1.3 µm). As a consequence, this diagnostic is unable to probe the dense shock front, the post-shock and the radiative precursor region very close to the shock front. Hard X-ray backlighting, on the other hand, is almost transparent to the typical conditions of a radiative precursor (i.e. very small changes in mass density) and to the best of our knowledge, so far, no experiment has been able to provide images of the entire shock structure (i.e. precursor, shock front and post-shock). Due to its moderate absorption in the conditions of the post shock and of the precursor, soft X-ray/XUV probing is an ideal tool to probe for this purpose.

Alternatively, by using ultrashort X-ray sources betatron or Compton [29, 30], based on the Laser Wakefield acceleration (LWFA) [31], it would be possible to resolve the dynamics of fast-moving shock front. The intrinsic temporal resolution of this X-ray source is less than 10 fs. In addition, betatron source have a very small source size (~µm), with this feature, it is possible to get the phase contrast image [32, 33] which provides both absorption and phase information. To drive such X-ray source a very high peak power laser system required. However, at the moment these sources are not available in large energy laser facilities.

In this paper, we present details of the design and results of radiative shock experiments performed at the PALS laser facility. We used the PALS 10 mJ Zn XUV laser (λ = 21.2 nm equivalent to a photon energy of ~58 eV with a pulse duration of ι = 150 ps) [[19],[34]] for instantaneous imaging of a radiative shock propagating at a velocity of ~ 50 $kms^{-1}$ in a rectangular shock tube filled with low-density xenon (~1.5 $mgcm^{-3}$ equivalent to ~0.3 bar) . This is a technically challenging diagnostic as, despite being one of the most energetic XUV lasers in the world, it requires very high-quality optics to minimize any beam losses. Thus, key elements in the experimental design were the use of a half cavity for the laser, an XUV multilayer spherical mirror for imaging, and appropriate target windows that could withstand the gas pressure

*Target Design for Xuv Probing of Radiative Shock Experiments*

inside the targets while allowing for sufficient XUV transmission. Moreover, the generation of the X-ray laser at PALS requires ~ 1 kJ of IR laser beam. As a result, the remaining energy to drive the shock was relatively small, (~60 J), which imposed a careful optimization of the target.

## 2. SHOCK TUBE TARGETS
### 2.1. Piston design and optimization

The strength of radiative effects on a shock driven in a gas depends mainly on two parameters: the shock velocity and the initial gas density. To generate a radiative shock, a high-power laser is focused onto a solid foil that acts as an ablator by converting the laser energy into mechanical momentum, i.e. behaving like a piston pushing and compressing the gas. Initially, during the ablation process a shock is generated which propagates through the foil and then breaks out into the gas. For a given laser intensity and within a given range of gas pressures, the shock velocity in the gas is mainly determined by the material and the thickness of the piston. To optimize the shock velocity, the thickness of the piston must be large enough so that the rear surface of the piston remains in its solid phase after laser irradiation. If the piston is too thin, the laser easily ablates it leading to insufficient mass to act as a piston. On the other hand, if the piston is too thick, the remaining mass is too large and the piston velocity is reduced, leading to slower shock velocities and reduced radiative effects. Thus, a careful selection of piston materials and thicknesses is crucial for the design of radiative shocks experiments.

The goal is to optimize this piston in order to get the highest velocity in xenon while limiting the xenon preheating by the ablated piston material. For the experiments presented here, the piston was made of a layer of polystyrene (hereafter CH) followed by a thin layer of gold in contact with the gas filling the target (xenon). This gold layer acts as a radiation shield preventing any pre-heating of the xenon and the walls of the gas-cell by X-rays generated during the CH ablation. The gold thickness should be thick enough in order to block the keV emission from the ablated CH but not too much in order not to slow down the shock. A gold layer with a thickness of 0.5 microns was used due to its transmission of ~1% at a photon energy of 1 keV.

In order to achieve the suitable intensity required to launch a radiative shock (a few $10^{14}$ Wcm$^{-2}$), a spot size of ~350 μm diameter is required. For the experiments to be in a regime where it is possible to produce a radiative precursor having moderate strong gas-absorption for XUV probing, the gas pressure P should be between ~0.1 bar and 0.5 bar. Thus, for the present experiments the xenon pressure was set to 0.3 bar (i.e. an initial mass density of $1.5 \times 10^{-3}$ gcm$^{-3}$).

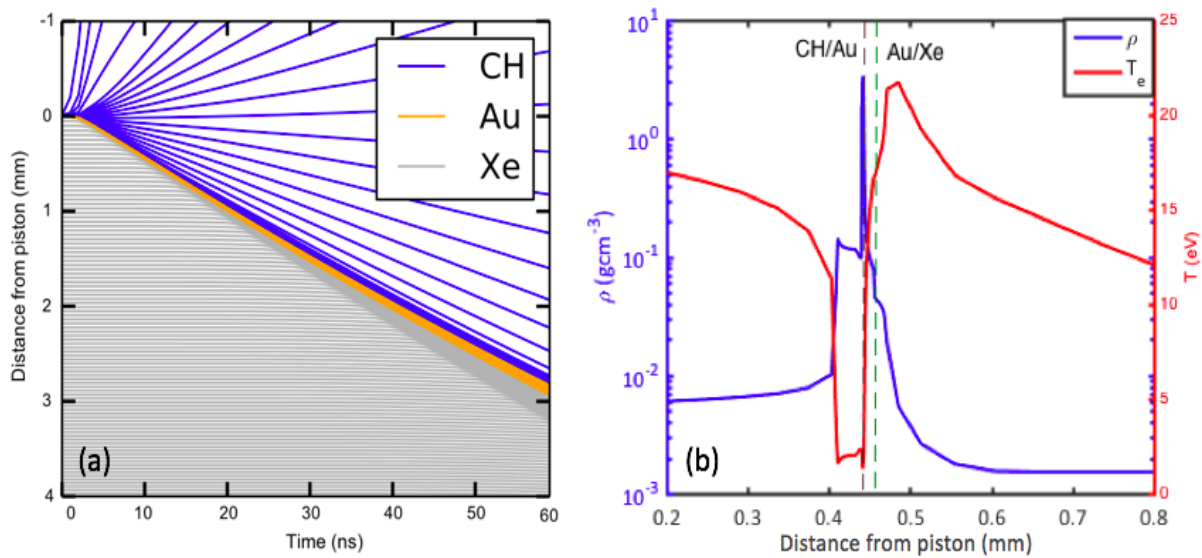

*Fig. 1: MULTI simulations in the case of a 10-μm piston made of CH followed by 0.5 μm Au layer and 6 mm of xenon gas at 0.3 bar. (a) Positions of the different layers versus time. The laser is incident from the top at t = 0 ns. (b) Mass density and electron temperature profiles as a function of distance from the piston at 10 ns after the arrival of the driving laser pulse. The laser is incident from the left and the zero position is the initial position of the piston.*

*U. Chaulagain, C. Stehlé, P. Barroso, M. Kozlova, J. Nejdl, F. Suzuki Vidal and J. Larour*

To optimize the design of the piston, the 1-D Lagrangian radiation hydrodynamics code MULTI [42] was used. The code solves the one-dimensional planar hydrodynamic equations coupled to the radiation transfer equations using a fully implicit numerical scheme. It uses tabulated equation of state (EoS) data, Planck and Rosseland opacities that are generated externally using a grey approximation. Thermal conduction is described within a flux-limited electron conduction model based on Spitzer's formula. The deposition of the laser is computed with an inverse-bremsstrahlung model. The simulation setup consisted of three layers: 10 μm CH, 0.5 μm Au, 6 mm of xenon at a 0.3 bar pressure and the laser parameters: $\lambda$ =1315 nm, a pulse duration of 350 ps and an energy of 60 J provides peak intensity of $1.3 \times 10^{14}$ Wcm$^{-2}$.

Fig 1a shows a position vs time plot of the different layer zones, with the 0 ns time marking the time when the driving laser hits the piston (incident from the top in the plot). The shock starts in xenon just after the end of the laser pulse. The CH layers move backwards due to ablation. The gold foil expands with time and the thickness of the post shock in xenon increases with time. From this figure, we can estimate the simulated velocity of the Xe/Au interface at about ~48 kms$^{-1}$. This velocity of the Xe/Au interface remains almost constant for more than 50 ns after laser pulse irradiation. The gold layer expands with time from 0.5 μm at t = 0 ns to 30 μm at = 20 ns.

The variation of mass density ($\rho$) and temperature (T) along the different material layers at 10 ns are reported in the Fig1b. From left to the right the profiles show a tail of dilute warm (~15 eV) CH layer at a density of 0.005 gcm$^{-3}$ (the initial density of CH was 1.1 gcm$^{-3}$), followed at ~0.4 mm by a dense shell of CH (up to ~0.03 gcm$^{-3}$) at a low temperature of ~2 eV. The gold layer is marked by a peak in mass density ~3 gcm$^{-3}$ (initial density of 19 gcm$^{-3}$, expands to a thickness up to 20 μm. Finally, the peak in electron temperature marks the position of the shock wave in Xenon which is located at ~0.48 mm.

In order to characterise the dependence of piston thickness for a fixed laser energy, simulations were ran for different thickness ranging between 5 and 25 μm of CH. The simulations show that the Au-Xe interface velocity for 25 μm CH is ~30 kms$^{-1}$ and for 5 μm CH is ~70 kms$^{-1}$ at 10 ns. For 5 microns, the velocity is at maximum. However, a close inspection of the mass density, temperature conditions indicate that most of the CH layer is ablated up to the gold layer, which in turn is heated more.

Thus 5 microns is the minimum thickness acceptable for the piston. In addition, thinner pistons are prone to technical difficulties related to target manufacture feasibility and high costs. Most importantly, thin pistons cannot withstand the pressure difference of 1 bar required during the gas filling process. Bearing these factors in mind and with the laser energy accessible to drive the shock at PALS, a 10 μm thick CH layer was chosen as the piston for the experiments. These 1D simulations were performed in order to optimize the thickness of the piston but not to compare the experiment. The experimental results were simulated using 2D Radiative Hydrodynamics code with the state-of-the-art opacities and the measured focal spot [20, 41] and laser energy reported in [46].

### 2.2. Shock tube design

Laser generated radiative shock targets can be divided broadly into two classes: shock tubes in which their internal cross section is adapted to the diameter of the shock wave, or cells with larger cross sections (see ref. [23, 24, 34, 37]. In the first case, the underlying idea is to produce a planar shock (i.e. with a 1-D geometry), facilitating the comparison with simulations and models. However, previous studies have shown that the radiation emitted by the shock can be absorbed by the lateral windows on the targets with losses due to, e.g. albedo effects, thus leading to a curved shock geometry instead of a planar one. In the second case, the large cell volume allows for an unimpeded propagation of the shock, leading to a spherical shock geometry (e.g. see ref. [27, 28, 38, 40]).

The targets in the experiments presented here are of the first type (shock tube), with a cross-section of $400 \times 400$ μm$^2$ and a length of 6 mm which, for a typical shock velocity of 50 kms$^{-1}$, allows to probe the shock up to ~ 100 ns. The feasibility of the target design was aided by INGESYM Engineering and the bulk of the targets (made of aluminium) were micro machined and then assembled at the Pole Instrumental de l'Observatoire de Paris. A customised target holder was designed to allow for repeatable target positioning and centering during each experiment with a precision within ~50 μm. The details of the target and its components are shown in Fig. 2.

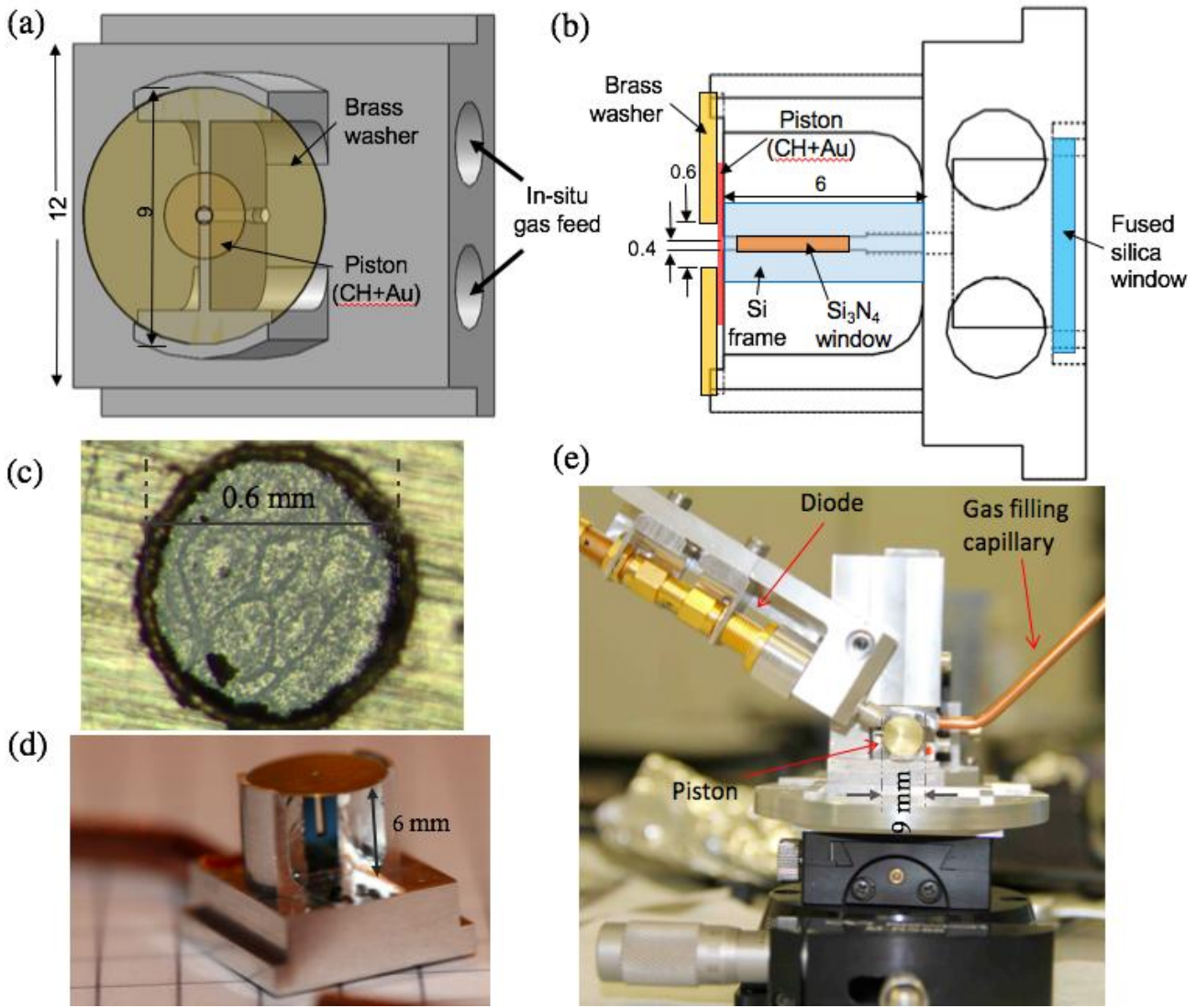

***Fig. 2 Target assembly:*** *(a) and (b) show front and side-on CAD views of the target, with the circular brass washer holding the CH piston shown schematically. (c) CH/Au piston seen from the side of polystyrene through the brass support (hole diameter 600 µm). (d) is the picture of the target and (e) target holder with a target and a high-speed IRD Silicon diode diagnostics.*

As previously mentioned, the shock tubes in the targets are closed at one end by a piston made of a 10 µm thick polystyrene (CH) foil coated with a 0.5 µm gold layer. The gold was deposited by thermal vacuum evaporation (Joule effect) with a precision estimated to 0.1 µm. The multilayer piston was glued to a 9-mm diameter, 0.1 mm thick brass washer with a 0.6 mm diameter aperture on it centre, shown in Figs. 2a-b. The brass disk was then glued to the aluminium target structure with a centering estimated to ±100 µm.

Fig. 2 (c) shows a microscope view of the Polystyrene side of the piston through the 0.6 mm aperture of the brass washer. Some defects can be seen the CH/Au foil (as striations with a width <~10 um) are not expected to influence the quality of the shock wave itself after the laser ablation process. However, it can't be guaranteed that they would not influence the development of instabilities when the shock wave generated by the interaction of the laser with CH propagates through the different layers.

The rear end of the tube was closed by a 500 µm thick fused silica window for target alignment before each shot. The alignment was done with a CW DPSS laser (532 nm, 40 mW) injected at this end through an optical fibre. Taking benefit of the tiny transmission of the piston and of a long exposure of a visible camera we were able to localise the channel and performed alignment adjustments. The target is filled inside the vacuum chamber with xenon at a pressure of 0.3 ± 0.01 bar through a rigid 1-mm copper capillary tube connected to the gas cell reservoir (see section 2.3).

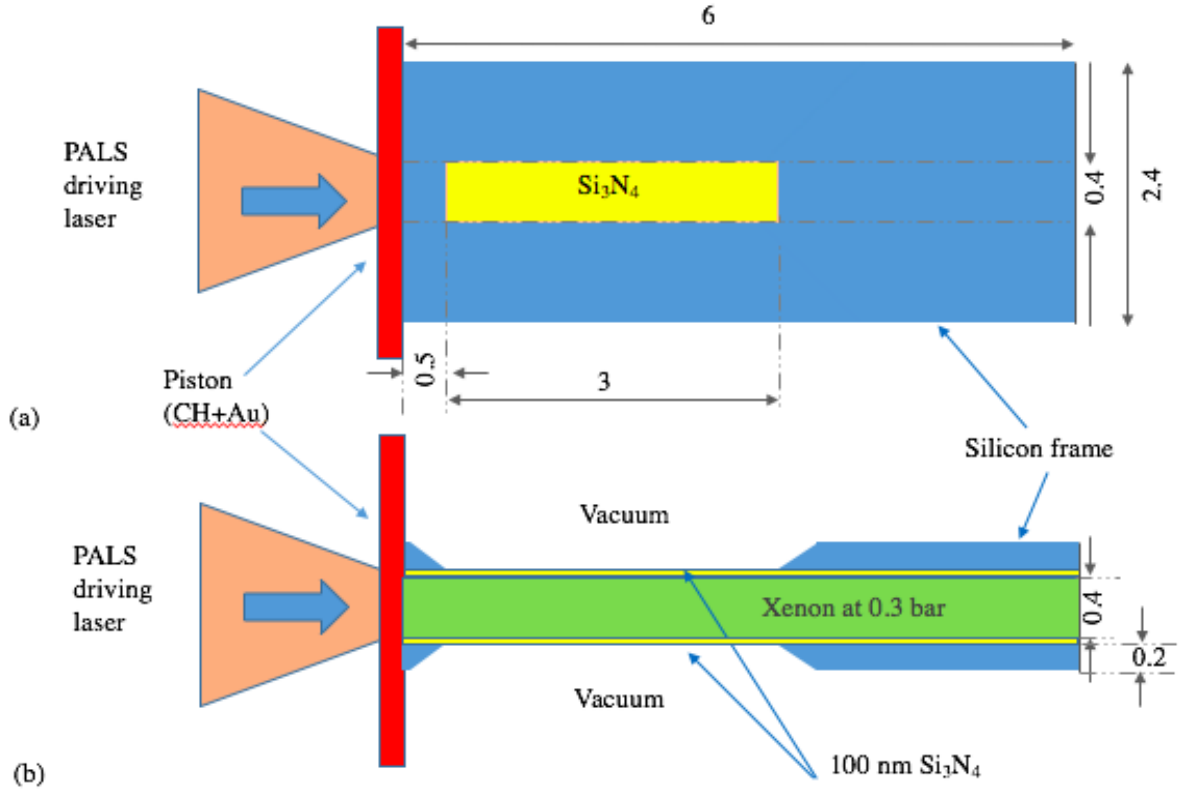

*Fig. 3: Schematic view of the lateral windows used for XUV probing of a radiative shock. The windows consist of a Si3N4 membrane supported by a silicon frame. The shock tube was closed from two sides laterally by 100 nm silicon nitride windows (size = 3 mm × 0.4 mm) which are supported on a 200 µm thick silicon frame (2.4 mm × 6 mm), a) side-on view, b) End-on cut view.*

In order to perform XUV probing of the plasma inside the targets, the lateral faces of the shock channel were enclosed by two 100 nm thick $Si_3N_4$ windows[1] (size = 3 mm × 0.4 mm) supported by a 200 µm thick Silicon frame with a dimension of 2.4 mm × 6 mm (see Fig. 3a and b)) and glued onto the Aluminium structure as shown as Fig. 2b and 2d. These windows are adapted for XUV imaging as well as for time-and-space resolved plasma self-emission measurements using high-speed XUV diodes. The selection of these $Si_3N_4$ windows was a critical step in the experimental design. The transmission of each of them at the probing wavelength (λ=21.2 nm) results in a net transmission through the 2 windows of 2%, which means a dramatic loss of the XUV laser energy. The important point is that they are able to withstand the working difference of a difference of pressure of up to ~ 0.3 -0.7 bar during the filling of the target inside the vacuum chamber before the experiments. These $Si_3N_4$ membranes were manufactured by SILSON using hydroxide-based anisotropic etching of silicon wafers coated with LPCVD silicon nitride [45]. The cut-view schematic view of the silicon nitride window in Fig. 3b shows the position of the silicon nitride membrane on the silicon frame. The silicon window was positioned in a way that the silicon

---

[1] The maximum pressure difference which is allowed for a membrane of thickness t, section a is given by:

$\Delta P = 2.44 \frac{t}{a}\sqrt{\frac{\sigma^3}{Y}}$ where $\sigma$ is breaking stress and $Y$ is the Young modulus. With $\sigma = 2 \times 10^9$ Nm$^{-2}$, $Y = 2.6 \times 10^{11}$ Nm$^{-2}$, t = 100 nm, and $a = 400$ µm of $Si_3N_4$ membrane, the pressure difference would be about 1 bar.

nitride membrane was facing the gas-side on the target in order to avoid the membranes detaching from the frame due to the pressure difference of 0.3 bar against vacuum inside the chamber.

The targets were then fixed to a compact target holder (Fig. 2(e)) which also supported a high-speed XUV diode for plasma self-emission measurement diagnostics allowing to monitor the arrival of the shock in the field of view of the diodes, and consequently to derive the shock velocity. The precise design of the targets and target holder resulted in a reproducibility better than ±50 μm in the target positioning before the shot.

### 2.3. Target filling and testing

The thin, 100 nm $Si_3N_4$ windows are prone to break for an uncontrolled large difference in pressure between the internal and external part of the cell. To limit the risks of breaking, an *in-situ* filling mechanism was designed. There is no reservoir; the pipe is directly connected to the filling mechanism. The system includes a piezoresistive pressure transducer (Swagelok PTI-S-AC.6-32), with a 10 kΩ bridge resistance working under a 14 - 30 V DC bias voltage. The theoretical response of the gauge was 0 V to 10 V for a range of absolute pressure: 0 to 1.6 bar, or relative pressure of −1 to + 0.6 bar. The target and vacuum chamber were pumped simultaneously, which was then followed by a slow (~1 minute) gas filling of the target. Once the set pressure was achieved, all valves were closed to isolate the target from the gas bottle and prevent any leakage of xenon into the chamber. A leakage test of the filling system was performed using a test target, filled at atmospheric pressure, and placed inside the vacuum chamber. The filling system required about 20 seconds to stabilise. The system was connected through a 20-m long coaxial cable to the control room allowing the remote reading of the pressure. A leakage test of the filling system was performed using a test target, filled at atmospheric pressure, and placed inside the vacuum chamber and observe the leakage, the leak rate was less than 5 mbar/min. All targets that were shot during the experiments were pre-tested for leaks to the working pressure of ~0.3 bar.

## 3. RESULTS AND DISCUSSION
### 3.1. PALS XUV Laser
The XUV laser was generated by irradiating a 3-cm-long high optically polished zinc slab by the PALS iodine laser (λ =1315 nm) using a double sequence with a controllable delay [23, 34, 39]. A pre-pulse (a few J) of the infra-red laser is used to produce a uniform pre-plasma, which is followed by a 500 J beam on target with a controlled delay of 10 ± 0.5 ns after the pre-pulse (for more details *see* [43]). The XUV laser is characterized by a wavelength of 21.2 nm, pulse duration of 150 ps, energy up to 4 mJ and peak spectral brightness of $10^{27}$ photons $s^{-1}mm^{-2}mrad^{-2}$. The beam has a divergence of 5.8 × 3.8 $mrad^2$. The coherence properties and high brightness of this laser make it well-suited for XUV imaging of dense plasmas [39]. This laser is used here for side-on XUV characterisation of the radiative shock.

### 3.2. XUV imaging
As already mentioned, XUV probing of dense radiative shocks in gases requires a bright source but also with a uniform energy distribution. The energy distribution of the Zn PALS XUV laser is uniform over a several $mm^2$ area [43]. For these experiments, and in addition to the $Si_3N_4$ windows, a critical element is the imaging XUV mirror which must have a uniform and high reflexivity at the wavelength of the probing laser. For this purpose, we used a periodic multilayer one-inch spherical mirror, with a substrate made of fused silica substrate (roughness 0.2 nm rms, pitch polished at Institut d'Optique) and with a deposition layers consisting of 20 periods of Al/Mo/$B_4$C with $B_4$C capping layer, made at CEMOX platform in Orsay [35, 44]. The reflectivity of this mirror is about 45% (at 21.2 nm) with a focal length of 300 mm.

During the experiments, the XUV laser is generated in a separate chamber from the target and reaches the xenon tube 20 ns after the driving beam (details of the experimental setup is reported in [20]). The laser beam passes side-on through the shock tube reaching the spherical multilayer mirror, which focuses the beam into an iron pinhole (diameter 500 μm) located at the focal plane of the spherical mirror. This pinhole acts as a spatial filter, reducing the stray light from the target, and thus improving the contrast on the CCD detector over the strong plasma self-emission [39]. Since the size of the pinhole is very small compared to the focal length of the spherical mirror, the aberrations linked to the mirror are kept at a very low level. After passing through the pinhole, the laser beam reaches the cooled CCD camera filtered with a 0.4 μm thick aluminium foil, which blocks the parasitic visible light coming from the chamber. The glass on the

*U. Chaulagain, C. Stehlé, P. Barroso, M. Kozlova, J. Nejdl, F. Suzuki Vidal and J. Larour*

vacuum cooled CCD camera (MTE 2048, Princeton Instrument) was removed to allow XUV transmission. The XUV beam is very bright, including two 100 nm $Si_3N_4$ membranes and 0.4 µm Al foil, in the imaging scheme, we have received over 25000 counts on the X-ray CCD. The magnification of the imaging setup was measured by placing a Nickel grid (G2002N, grid pitch =125 µm from Agar Scientific Ltd) on a test target. The section of the grid viewed by the probing laser is shown in the (a-b). Taking the pitch of the grid of 125 µm, the magnification of the imaging setup is measured to be equal to 4.85. The limiting spatial resolution of the imaging setup is about 22 µm in the object coordinate.

A sample snapshot of XUV image of a radiative shock is reported in the Fig. 4 (c). The details explanation of this shock image and its analysis has been already published Ref [20]. In Fig. 4 (c), we have reported the XUV image of the radiative shock section viewed through the Si3N4 windows at 20 ns after shock the lunching of the shock. The XRL image shows three different regimes of the radiative shock: the post-shock, shock front and the radiative precursor. The transmission at 21.2 nm along a longitudinal section of the image is along a line (yellow) at the position y = 250 µm together with the transmission at 25 µm below and above this line is reported in the Fig. 4d. A strong absorption of the probing laser by the piston residues (CH and gold) is seen on the left end of the image with dark zone (x < 680 µm) with a transmission of about 5 - 15%. The position of shock front can be identified by rapid decrease of the transmission between 980 and 1020 µm with the XRL transmission of value of 35%. The shock is curved and elongated with its front located at about 1000 µm. The extension of radiative precursor is up to 1180 µm with a transmission of up to 85%. The thickness of the post-shock in the image is about 30 µm. The position of shock front at 1000 µm, with a rapid drop in transmission due to strong absorption, which gives the shock front velocity is about 50 $kms^{-1}$.

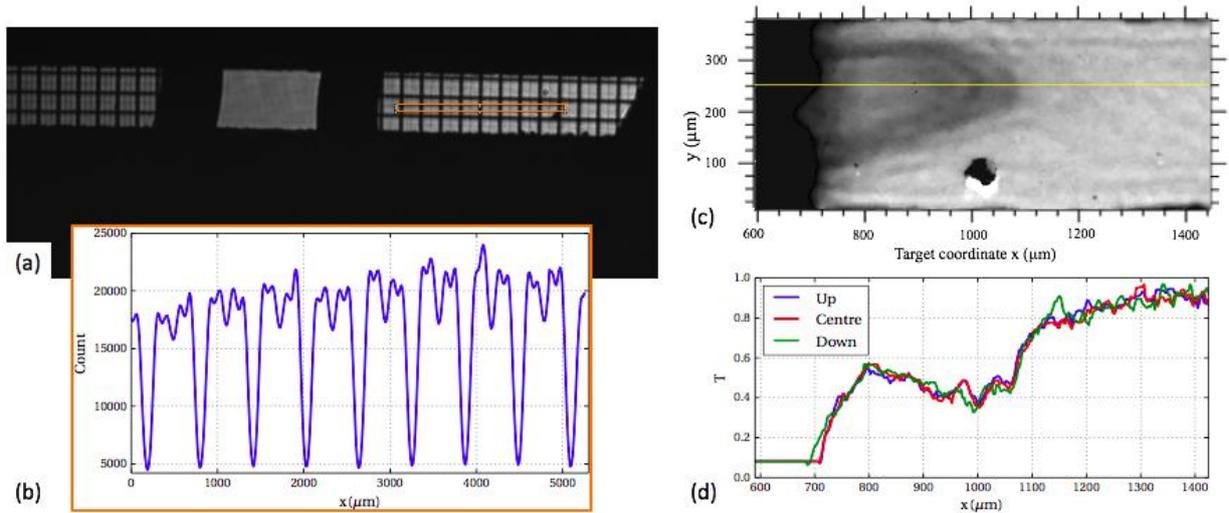

**Fig. 4:** *(a) Magnification of imaging setup calculated by placing a grid a grid placed side-on on a test target. (b) Line out of the selected zone of the grid and the x-axis with the image coordinates (c) Crop of the section highlighting the shock feature. (d) transmission profiles along a horizontal section at the position of the yellow line on the top and 25 µm above and below of it. The profiles are normalized to the maximum signal in the image which is at the right end of the image where the gas is unperturbed.*

This result is consistent with the plasma self-emission measurement using the high-speed XUV diode reported in the Ref [20] and also supported by the 2D numerical simulations obtained with ARWEN code [46]. The second bump at ~ 1070 µm has been attributed to the behaviour of the xenon opacity at 21.2 nm for a given density and the temperature of the radiative precursor [48]. The state of art monochromatic opacities of xenon at the XUV probing wavelength for different temperature of shock region were calculated using the RAPCAL code [47], the resulted are reported in the Ref [20].

## 4. CONCLUSIONS

The novel design of the target together with the high energy Zn XUV laser at the PALS laser facility we were able to show, for the first time, an unambiguous and complete image of the pre- and post-shock regions in a radiative shock. These preliminary experiments open the door for future investigations of such hypersonic flows. Among possible improvements it is worth pointing out: 1) the transmission of the XUV target windows can be improved by using Silicon windows instead of $Si_3N_4$ windows due to their increased transmission of 65% per window for a 150 nm thickness compared to a 13% transmission for 100 nm $Si_3N_4$. Preliminary tests of silicon windows with a thickness of 150 nm have been performed at PALS showing promising results. 2) The uniformity of the CH/Au piston can be improved by using Parylene-N instead of polystyrene [26]. Further improvements in the XUV imaging mirrors and in the energy of PALS's XUV laser would allow to probe denser plasma conditions in a radiative shock by implementing e.g. XUV interferometry and temperature measurements with X-ray Thomson scattering. Current X-ray backlighting sources have large source sizes (~ 100 μm) and longer pulses which are not achieving the required resolution to characterize dense plasmas. However new ultrashort X-ray sources based on LWFA that can be scale from few keV to 10's of keV [49], have been considered to be implemented in the large-scale laser system [50-52]. These sources can provide a high temporal resolution and high contrast image of fast-moving shock waves and enhance the diagnostics capability to study the dynamics of the dense plasmas.

**ACKNOWLEDGEMENTS**

The authors would like to thank Pascal Jagourel, Florent Reix, Thierry Melse from Pole Instrumental de l'Observatoire de Paris for their help in the target design. We acknowledge the support of the Scientific Council of Observatoire de Paris, the French CNRS/INSU Program PNPS, the Labex PLAS@PAR (ANR-11-IDEX-0004-02), the UFR de Physique of UPMC as also of EXTREME LIGHT INFRASTRUCTURE project no. CZ.1.05/1.1.00/02.0061,CZ.02.1.01/0.0/0.0/15_008/0000162 and EC OP projects no. CZ.1.07/2.3.00/30.0057 and CZ.1.07/2.3.00/20.0279, which were co-financed by the European Social Funds and the state budget of the Czech Republic and LASERLAB-EUROPE (grant agreement no. 284464, ECs Seventh Framework Programme). U.C, M. K. and J.N. would like to acknowledge the project Advanced research using high intensity laser produced photons and particles (CZ.02.1.01/0.0/0.0/16_019/0000789) from European Regional Development Fund (ADONIS). The results of the Project LQ1606 were obtained with the financial support of the Ministry of Education, Youth and Sports as part of targeted support from the National Programme of Sustainability II. We acknowledge project no. CZ.1.07/2.3.00/20.0279 that was co-financed by the European Social Fund and the state budget of the Czech Republic.